\documentclass[12pt]{article}
\pdfoutput=1
\usepackage{amsfonts}
\usepackage{url}
\usepackage{hyperref}
\usepackage{bbold}
\usepackage{slashed}
\usepackage{graphicx}
\usepackage{pgfplots}
\usepackage{color}
\usepackage{mathrsfs}
\usepackage{fancybox}% http://ctan.org/pkg/fancybox
\usepackage{lipsum}% http://ctan.org/pkg/fancybox
\usepackage{eurosym}
\usepackage{tcolorbox}
\usepackage{lmodern}
\usepackage{tikz}
\usepackage{tikz,pgfplots}
\usepackage{pgfplots}
\usepackage{amsmath,amsthm,amssymb}
\usepgfplotslibrary{fillbetween}
\usepackage{feynmp}
\usepackage{slashed}
\usepackage{graphicx}
\usepackage{epstopdf}
\usepackage{subfigure}
\usepackage{color}
\usepackage{pgfplots}
\usetikzlibrary{decorations.pathmorphing}
\tikzset{snake it/.style={decorate, decoration=snake}}
\usetikzlibrary{shapes}
\usepackage{empheq}
\usepackage{mathrsfs}

\def\({\left (}
\def\){\right )}
\def\[{\left [}
\def\]{\right ]}
\def\d{\mathrm{d}}

\numberwithin{equation}{section}

\usepackage{environ}
\NewEnviron{myequation}{%
\begin{eqnarray}
\scalebox{1.1}{$\BODY$}
\end{eqnarray}
}

\newcommand{\beq}{\begin{equation}}
\newcommand{\eeq}{\end{equation}}
\newcommand{\nn}{\nonumber\\} 
\newcommand{\bea}{\begin{eqnarray}}
\newcommand{\ea}{\end{eqnarray}}
\newcommand{\barr}{\!\begin{array}}
\newcommand{\earr}{\end{array}\!}
\newcommand{\lb}{{\langle}}
\newcommand{\rb}{{\rangle}}

\def\d{{\partial}}
\def\n{{\bf \widehat n}}
\def\k{{\bf k}}

\begin{document}
\begin{titlepage}

\setcounter{page}{1} \baselineskip=15.5pt \thispagestyle{empty}

\vfil

${}$
\vspace{1cm}

\begin{center}

\def\thefootnote{\fnsymbol{footnote}}
\begin{changemargin}{0.05cm}{0.05cm} 
\begin{center}
{\Large \bf  An Inelastic Bound on Chaos}
\end{center} 
\end{changemargin}

~\\[1cm]
{Gustavo J. Turiaci\footnote{\href{mailto:turiaci@ucsb.edu}{\protect\path{turiaci@ucsb.edu}}}}
\\[0.3cm]

{\normalsize { \sl Physics Department,  %and ${}^{\rm b}$Princeton Center for Theoretical Science 
%Princeton University Princeton, NJ 08544, USA\\[1.0mm]
University of California, Santa Barbara, CA  93106, USA}} \\[3mm]

\end{center}

%\vspace{2cm}
%\vspace{1cm}

%\hrule
 \vspace{0.2cm}
\begin{changemargin}{01cm}{1cm} 
{\small  \noindent 
\begin{center} 
\textbf{Abstract}
\end{center} }
We study a generalization of the chaos bound that applies to out-of-time-ordered correlators between four different operators. We prove this bound under the same assumptions that apply for the usual chaos bound and extend it to non-hermitian operators. In a holographic theory, these correlators are controlled by inelastic scattering in the bulk and we comment on implications. In particular, for holographic theories the bound together with the equivalence principle suggests that gravity is the highest spin force, and the strongest one with that spin. 

\end{changemargin}
 \vspace{0.3cm}
%\hrule
\vfil
\begin{flushleft}
\today
%March 20, 2013
\end{flushleft}

\end{titlepage}

%\newpage
%\tableofcontents
%\newpage

\addtolength{\abovedisplayskip}{.5mm}
\addtolength{\belowdisplayskip}{.5mm}

\def\plus{\raisebox{.5pt}{\tiny$+$\smpc}}

\addtolength{\parskip}{.6mm}
\def\spc{\hspace{1pt}}

\def\nspc{{\hspace{-2pt}}}
\def\ff{\rm\smpc f\smpc} 
\def\fff{\mbox{Y}}
\def\ww{{\rm w}}
\def\smpc{{\hspace{.5pt}}}

\def\zz{{\spc \rm z}}
\def\xx{{\rm x\smpc}}
\def\xxi{\mbox{\footnotesize \spc $\xi$}}
\def\jj{{\rm j}}
 \addtolength{\baselineskip}{-.1mm}

\renewcommand{\Large}{\large}

\def\calO{{b}}
\def\be{\begin{equation}}
\def\ee{\end{equation}}

%\begin{document}

%\keywords{}


\def\mathbi#1{\textbf{\em #1}} 
\def\som{{ \textit{\textbf s}}} 
\def\tom{{ \textit{\textbf t}}} 
\def\nom{n} %{{ \textit{\textbf n}}} %{\mbox{\fontsize{13pt}{.15pt}${ \smpc \mathbf{n}\smpc }$}}}
\def\mom{m} %{ \textit{\tex%mbox{\fontsize{9.5pt}{.1pt}$\mathbf m$}}}
%\def\kom{{\textbf{\em k}}}
%_{{}%_{\mbox{\fontsize{4pt}{.5pt}{\omega}}}}}
%\def\mom{{\textbf{\em m}}}
%_{{}_{\mathtt{\omega'}}}}}
\def\la{\langle}
\def\bea{\begin{eqnarray}}
\def\eea{\end{eqnarray}}
\def\is{\!  & \!  = \!  &  \!}
\def\ra{\rangle}
\def\half{{\textstyle{\frac 12}}}
\def\cL{{\cal L}}
\def\halfi{{\textstyle{\frac i 2}}}
\def\ba{\bea}
\def\ea{\eea}
\def\lb{\langle}
\def\rb{\rangle}
\newcommand{\rep}[1]{\mathbf{#1}}

\def\uU{\bf U}
\def\be{\bea}
\def\ee{\eea}
\def\delbar{\overline{\partial}}
\def\ra{\rangle}
\def\la{\langle}
\def\ccdot{\!\spc\cdot\!\spc}
\def\nspc{\!\spc\smpc}
\def\tr{{\rm tr}}
\def\li{|\spc}
\def\ri{|\spc}

\def\hf{\textstyle \frac 1 2}

\def\bfcdot{\raisebox{-1.5pt}{\bf \LARGE $\spc \cdot\spc $}}
\def\spc{\hspace{1pt}}
\def\is{\! &  \! = \! & \!}
%Symbols
\def\d{{\partial}}
\def\n{{\bf \widehat n}}
\def\k{{\bf k}}
\def\GO{{\cal O}}

\def\pp{{\mbox{\tiny$+$}}}
\def\mm{{\mbox{\tiny$-$}}}

\setcounter{tocdepth}{2}
%\tableofcontents
\addtolength{\baselineskip}{0mm}
\addtolength{\parskip}{.6mm}
\addtolength{\abovedisplayskip}{1mm}
\addtolength{\belowdisplayskip}{1mm}

\def\sss{{\bf s}}

\def\fff{e}
\def\lL{{{}_{{}^{L}}}}

\def\rR{{{}_{{}^{R}}}}

\def\llL{{{\! \spc}_{{}^{L}}}\!\spc}
\def\rrR{{{\!\spc}_{{}^{R}}}\!\spc}

\section{Introduction}
 The study of high energy scattering near the horizon of a black hole has lead to an improved perspective on quantum chaos \cite{LO, Dray:1984ha, SS, KitaevTalks, MSS}. The scrambling of information near the horizon of the black hole is related to the chaotic spread of information on the boundary quantum system. 
 
 The signature of quantum chaos used in this context is given by the exponential growth of the square of double commutators $\lb |[A(0),B(t)]|^2 \rb_\beta$ evaluated at a thermal state with inverse temperature $\beta$,  \cite{ KitaevTalks,LO}. The main object needed to compute this double commutator is the real part of the out-of-time-ordered (OTOC) correlator $\lb A^\dagger(0) B^\dagger(t) A(0) B(t) \rb_\beta$. The connected part of the OTOC grows exponentially with time in a chaotic system, with a rate defined as the Lyapunov exponent $\lambda$. This growth happens long after the dissipation time $t_d$ controlled by the thermalization scale of time-ordered correlators. At a much larger scrambling time $t_{\rm sc}$ the connected part of the OTOC, and also the commutator, saturate.
 
 In reference \cite{SS} it was explained in detail how, in holographic theories, the bulk computation of an OTOC is given by a high energy scattering near the black hole horizon \cite{Dray:1984ha}. The result is a convolution between wavefunctions (bulk-boundary propagators), that evolve the particles from the boundary to the near horizon region, and a local high energy S-matrix near the horizon. A high energy gravitational scattering is equivalent to a classical shockwave interaction \cite{Dray:1984ha}. This gives a Lyapunov exponent $\lambda = 2 \pi / \beta$ which was shown in \cite{MSS} to be maximal. When doing this calculation in string theory, and assuming inelastic effects are subleading, reference \cite{SS}, building upon \cite{Brower:2006ea}, explains how the sum over all stringy modes are equivalent to an effective elastic Pomeron which produces a perturbative correction $\lambda = \frac{2\pi }{\beta} (1 - \mathcal{O}(\alpha'))$, where $\alpha'$ is related to the string tension, which is analogous to the flat space Regge asymptotics. On the other hand, the saturation of the OTOC at the scrambling time is related to the decay of the bulk-boundary propagators and therefore to the quasinormal modes of the black hole.
 
So far the attention has focused on OTOC that appear on double commutators, since they are more directly related to the definition of chaos of \cite{LO} as explained above. In this note we will analyze general OTOC between four arbitrary operators. For holographic theories, the importance of these quantities is more evident in the bulk. When the operators are different, the bulk scattering is completely inelastic and the Pomeron controlling these OTOC does not necessarily have the quantum numbers of the vacuum, for example. In particular, gravity plays no role in the local near horizon bulk interaction, and the growing piece of these correlators probe interactions that are not universal. Therefore it is interesting to study them to get more information about the bulk. 

In this paper we will extend the chaos bound and constrain arbitrary out-of-time ordered correlators. The assumptions we will use are the same as in the original chaos bound of \cite{MSS}. In particular, we will focus for simplicity on hermitian operators although we will relax this assumption later. For the exponential ansatz of an OTOC we will use the notation
\bea\label{intro:eq:}
{\rm Re}~{\rm Tr} [ y A(0) y B(t) y C(0) y D(t) ] \approx F_d -  \varepsilon_{ABCD} e^{\lambda_{ABCD} t},
\ea
where by $F_d$ we denote the order one factorized approximation (which implicitly also depends on the choice of operators) and $\varepsilon$ is a small correction. Following \cite{MSS} we define $y$ such that $y^4=e^{-\beta H}/Z$.   For systems with a large number of degrees of freedom $N$ the amplitude of the growing piece is of order $\varepsilon\sim N^{-2}$ while the factorized piece is generically of order one. Unless all four operators $A$, $B$, $C$ and $D$ are all different, the correlator in \eqref{intro:eq:} is real. Correlators involving combinations $ABAB$, $ABCB$ or $ABAD$ are all real. 

We will refer to configurations such as $ABAB$ that appear on double commutators as `diagonal' or `elastic' OTOC, while we will refer to the generic OTOC as `off-diagonal' or `inelastic' correlators.

In this notation, the chaos bound of \cite{MSS} is a statement about the positivity of the prefactor $\varepsilon_{ABAB}>0$ and a bound on the growth rate $\lambda_{ABAB} \leq 2\pi/\beta$. This is valid for any choice of $A$ and $B$, even though in most examples the Lyapunov exponent is independent on the choice of operators. We will show (see details in section \ref{sec:proof}) that the inelastic OTOC for four different operators is also similarly constrianed
\beq
\lambda_{ABCD} \leq \lambda_{\rm diag} \leq \frac{2\pi}{\beta},
\eeq
where $\lambda_{\rm diag} \equiv {\rm min} (\lambda_{ABAB},\lambda_{ADAD},\lambda_{CBCB}, \lambda_{CDCD})$. This means a generic off-diagonal OTOC cannot grow faster than diagonal ones. From the gravity side, this puts a bound on the spin of the effective mode controlling this interaction, it cannot be bigger than $2$. 

It is reasonable to expect all diagonal or off-diagonal OTOC for arbitrary $A$, $B$, $C$, $D$ to grow with the same rate $\lambda_L$, even if not maximal \footnote{For example in the SYK model \cite{Sachdev:1992fk, KitaevTalks,Maldacena:2016hyu,Kitaev:2017awl,Gu:2018jsv} the exponentially growing piece always comes from the same set of ladder diagrams, regardless of how we glue it to external operators.}. With this assumption, we can also bound the amplitude of the growing piece. In the general case of four different operators the constraint is presented in section \ref{sec:proof}. If we take two of the operators to be the same then we can write a simpler version  
\beq\label{ec:introbound}
(\varepsilon_{ABCB})^2 \leq \varepsilon_{ABAB} \varepsilon_{CBCB},
\eeq
and similarly for $\varepsilon_{ABAD}$. The same structure is present for the case of an OTOC between four different operators, $\varepsilon_{ABCD}$  is bounded by the prefactors appearing in diagonal correlators. From the gravity side this puts a bound on how strongly can matter couple to the effective mode controlling this interaction. 

In the context of holographic theories, the coefficient in the left hand side of the inequality \eqref{ec:introbound} is given by an inelastic scattering between particles in the bulk, which in general does not involve graviton exchange. It is interesting to see that we can bound such a process by the right hand side, which is universally fixed by gravitational interactions and the equivalence principle. Even though the inelastic coupling $\varepsilon_{ABCD}$ does not necessarily have a definite sign, its magnitude cannot be bigger than a mean of the gravitational couplings. In this context, this analysis suggests that gravity is the highest spin interaction, and the strongest with that spin.

In section \ref{sec:nonh} we generalize the chaos bound to non-hermitian operators. Then similar constraints on an OTOC between four non-hermitian operators can be derived.

In section \ref{sec:2dCFT} we make some comments regarding the behavior of inelastic OTOC for 2d CFT.  In \cite{Roberts:2014ifa} the authors show how the maximal chaos exponent is controlled by the dominance of the identity Virasoro block. For an OTOC between different operators the identity channel does not appear in the OPE expansion. Using semiclassical expressions for non-vacuum blocks at large central charge $c$ we study the behavior of off-diagonal OTOC. In particular, we see that before the scrambling time, Virasoro descendants are not important in the second sheet. This shows how gravity naively plays no role in the physics of these OTOCs. After the scrambling time at which exponential growth stops, we show how quasinormal modes dictate the decay of the OTOC. 

We conclude in section \ref{Sec:conc} with open questions and future directions.

\section{Constraints on generic OTOC}\label{sec:proof}

In this section we will show how to bound general OTOC between arbitrary operators. The argument is simple but requires some notation. In order to do that, we will begin by stating the chaos bound from \cite{MSS}, which we will refer to as the elastic chaos bound. 

In \cite{MSS} the authors focus on a particular correlator 
\beq
F(t) \equiv {\rm Tr} [ y A(0) y B(t) y A(0) y B(t) ],
\eeq
between hermitian operators $A$ and $B$, where divergences are regularized by placing them symmetrically along the euclidean circle of size $\beta$. This is implemented by inserting the operators $y$ defined as $y^4=e^{-\beta H}/Z$. The motivation for considering such correlators comes from its relation to commutators square between operators $A(0)$ and $B(t)$. 

We will consider times that are much larger than the dissipation time $t_d$ but smaller than the scrambling time $t_{sc}$, which we assume to be parametrically larger (as in, for example, large $N$ theories). In this regime the OTOC is almost constant and given by its factorized contribution $F_d$ to leading order, where 
\beq
F_d \approx {\rm Tr} [ y^2 A y^2 A] {\rm Tr} [ y^2 B y^2 B] 
\eeq
For chaotic systems we expect the subleading behavior to have an exponential behavior 
\beq\label{eq:ans}
F(t) = F_d - \varepsilon ~e^{\lambda t} +\ldots 
\eeq
where $\lambda$ is the Lyapunov exponent of the system. The parameter $\varepsilon$ is a small constant which controls the scrambling time at which the OTOC decays. For a large $N$ system it is of order $\varepsilon \sim N^{-2}$. From now on we will denote these prefactors of exponentially growing terms by $\varepsilon$ to denote they are small compared to the factorized term. 

The chaos bound from \cite{MSS} states that the quantity $F(t)$ is bounded by the right hand side of equation \eqref{eq:ans} with both 
\beq
\varepsilon \geq 0 ~~~{\rm and}~~~\lambda \leq \frac{2\pi}{\beta}
\eeq
We will take this as our starting point for the generalizations below. Therefore we will implicitly use the same assumptions and caveats as in \cite{MSS}.

\subsection{An inelastic chaos bound}

In this section we will prove the bound stated in the introduction regarding OTOC between four different operators. We will focus first on hermitian operators. The upshot is that the growing piece of a general OTOC  cannot grow faster than exponentially, with the maximal rate $\lambda = 2\pi/\beta$. We will also see how to put a bound on the magnitude of the growing piece.

To simplify the presentation, we will go over the argument in steps. We will first generalize the chaos bound to a correlator ${\rm Tr} [ y A(0) y B(t) y C(0) y B(t) ]$. This OTOC is real for arbitrary operators since 
\beq\label{eq:realABCB}
{\rm Tr} [ y A y B(t) y C y B(t) ]^\dagger = {\rm Tr} [ B(t) y C y B(t) y A y ]={\rm Tr} [ y A y B(t) y C y B(t) ],
\eeq
where we used that the operators are hermitian. In the first line we applied the hermitian conjugation inside the trace and in the second one used the cyclic property of the trace\footnote{Similarly, one can show that $ {\rm Tr} [ y A y B(t) y A y D(t) ]$ is real and symmetric under exchange of $B$ and $D$. }. Moreover, the OTOC is also symmetric under the exchange of $A$ and $C$ 
\beq\label{eq:symABCB}
{\rm Tr} [ y A y B(t) y C y B(t) ]={\rm Tr} [ y C y B(t) y A y B(t) ]
\eeq
which follows from the cyclic property of the trace.

To bound ${\rm Tr} [ y A(0) y B(t) y C(0) y B(t) ]$ we will analyze a diagonal correlator of the form
\bea\label{eq:corrvwvw}
F(t)= {\rm Tr} [ y V y B(t) y V y B(t) ]  ,~~~ V=\alpha_1 A + \alpha_2 C,
\ea
for arbitrary real coefficients $\alpha_1$ and $\alpha_2$. To simplify the notation, we omit the time argument when the operator is inserted at $t=0$. Expanding each term in the right hand side of equation \eqref{eq:corrvwvw} gives 
\bea\label{eq:ABCBexpand}
F(t)  &=& \alpha_1^2  {\rm Tr} [ y A y B(t) y A y B(t) ]+\alpha_2^2  {\rm Tr} [ y C y B(t) y C y B(t) ] \nn
&& + 2 \alpha_1 \alpha_2 {\rm Tr} [ y A y B(t) y C y B(t) ].
\ea
This contains the correlator we want to bound. There for by using the information we learn from the chaos bound on diagonal OTOC we can bound off-diagonal OTOC such as ${\rm Tr} [ y A y B(t) y C y B(t) ]$.

Before we move on we can write an ansatz for these OTOC similar to equation \eqref{eq:ans}. For concretenes and to set notation we write 
\bea
{\rm Tr} [ y A y B(t) y A y B(t) ] &=& F^{AA}_d - \varepsilon_{AA}~ e^{\lambda_{AA} t},\\
{\rm Tr} [ y C y B(t) y C y B(t) ] &=& F^{CC}_d-\varepsilon_{CC} ~e^{\lambda_{CC} t},\\
{\rm Tr} [ y A y B(t) y C y B(t) ] &=&F^{AC}_d- \varepsilon_{AC} ~e^{\lambda_{AC} t}.
\ea
Where we indicate the dependence on the operators of the factorized leading contribution $F_d$, the amplitude of growing piece $\varepsilon$ and rate $\lambda$. We leave the dependence on the operator $B$ implicit. From \eqref{eq:realABCB} and \eqref{eq:symABCB} we know that $F_d^{AC}=F_d^{CA}$, $\varepsilon_{AC}=\varepsilon_{CA}$ and $\lambda_{AC}=\lambda_{CA}$ are real. 

To leading order, the right hand side of equation \eqref{eq:ABCBexpand} above is approximately constant in time, order one, and given by
\beq
F_d \approx \alpha_1^2 F_d^{AA} + \alpha_2^2 F_d^{CC} + 2 \alpha_1 \alpha_2 F_d^{AC}.
\eeq
This quantity is positive for any choice of $\alpha_1$ and $\alpha_2$. This can be shown using Cauchy-Schwarz or more directly by starting from expression \eqref{eq:corrvwvw} in terms of $V$. Moreover, we might also diagonalize the $2\times 2$ matrix of two-point functions between $A$ and $C$ such that $F_d^{AC}=0$, making the equation above manifestly positive.

Next, we will focus on the subleading piece growing in time. We will consider first the most general case where all rates exponents are allowed to be different. Using this ansatz for the maximal growth we can write the subleading part of the OTOC as
\beq\label{eq:ABCBsublead}
F_d - F(t)= \alpha_1^2 \varepsilon_{AA} e^{\lambda_{AA} t} + \alpha_2^2 \varepsilon_{CC}  e^{\lambda_{CC} t} + 2 \alpha_1 \alpha_2 \varepsilon_{AC} e^{\lambda_{AC} t}.
\eeq
 Since $\alpha_1$ and $\alpha_2$ are arbitrary coefficients, and by the elastic chaos bound, we can conclude that $\lambda_{AA}, \lambda_{CC}, \lambda_{AC} \leq 2\pi/\beta$. Otherwise we could form a linear combination of $A$ and $C$ such that $F(t)$ could violate the chaos bound. Moreover we can also argue that $\lambda_{AC} \leq {\rm min}(\lambda_{AA},\lambda_{CC})$. Otherwise eventually the mixed term would dominate and we could choose a sign of the $\alpha$'s which would give a negative prefactor, also violating the chaos bound. In other words, $\lambda_{AC} > {\rm min}(\lambda_{AA},\lambda_{CC})$ would imply $\varepsilon_{AC}=0$. 
 
Above we considered the most general case. Now we will assume that all OTOC grow with the same rate $\lambda$. In this case the chaos bound on the prefactor sign gives  us a bound
\beq
\alpha_1^2 \varepsilon_{AA} + \alpha_2^2 \varepsilon_{CC}  + 2 \alpha_1 \alpha_2 \varepsilon_{AC} \geq 0, ~~~~\forall ~\alpha_1,\alpha_2
\eeq
coming from the diagonal chaos bound applied to the right hand side of equation \eqref{eq:ABCBsublead}. This condition is equivalent to the following constraint 
\beq\label{eq:boundabac}
\varepsilon_{AC}^2 \leq \varepsilon_{AA} \varepsilon_{CC}
\eeq
From this condition we see that even though we can constrain the growth of $\lb A B CB \rb$, the chaos bound does not constrain the sign of the correction, which could be positive or negative but with a magnitude bounded by $\sqrt{\varepsilon_{AA} \varepsilon_{CC}}$. This is also analogous to the ANEC case studied in \cite{CMT} (see also \cite{Meltzer:2017rtf}).

Having done this, the obvious next step is to consider other linear combinations. An option is  
\beq\label{eq:ABCDlc}
F(t)= {\rm Tr} [ y A y W(t) y A y W(t) ]  ,~~~ W=\alpha_1 B + \alpha_2 D.
\eeq
The chaos bound applied to this correlator gives analogous bounds as the previous analysis for the (real) correlator $ {\rm Tr} [ y A y B(t) y A y D(t) ]$. Namely, the growing piece cannot grow too fast and the amplitude cannot be bigger than diagonal one. Instead, to obtain new bounds, we will consider
\beq\label{eq:ABCDlc}
F(t)= {\rm Tr} [ y A y W(t) y C y W(t) ]  ,~~~ W=\alpha_1 B + \alpha_2 D
\eeq
with real coefficients $\alpha_1$ and $\alpha_2$. Then we can use the inelastic chaos bound derive above to constrain ${\rm Tr} [ y A y B(t) y C y D(t) ]$. We again assume an exponential ansatz on each term. A new feature of the most general case is that the mixed term now is not real anymore since 
\beq
{\rm Tr} [ y A y B(t) y C y D(t) ]^\dagger = {\rm Tr} [ y A y D(t) y C y B(t) ] = {\rm Tr} [ y C y B(t) y A y D(t) ].
\eeq
This means that exchanging $A \leftrightarrow C$ or $B \leftrightarrow D$ are related by complex conjugation. Only a simultaneous exchange of $A \leftrightarrow C$ and $B \leftrightarrow D$ is a symmetry. From expanding the right hand side of \eqref{eq:ABCDlc} we see it is only sensitive to the real part of ${\rm Tr} [ y A y B(t) y C y D(t) ]$.

To set notation we write the exponential ansatz for the correlators as 
\bea
{\rm Tr} [ y A y B(t) y A y B(t) ] &=& F_d^{ABAB} - \varepsilon_{ABAB} ~e^{\lambda_{ABAB} t}, \\ 
{\rm Tr} [ y A y B(t) y C y B(t) ] &=& F_d^{ABCB} - \varepsilon_{ABCB} ~e^{\lambda_{ABCB} t},\\
{\rm Re}~{\rm Tr} [ y A y B(t) y C y D(t) ] &=& F_d^{ABCD} - \varepsilon_{ABCD} ~e^{\lambda_{ABCD} t}.
\ea
In these expressions all quantities on the right hand side are real. In the third line, after taking the real part, the quantities are symmetric under independently exchanging $A$ and $C$ or $B$ and $D$. 

Now we can expand the right hand side of \eqref{eq:ABCDlc}. Again, the factorized contributions $F_d$ give some leading constant piece for the correlator in \eqref{eq:ABCDlc} and we will focus on the subleading growing piece. From the constraint on the growth rate in time we obtain the following bound quoted in the introduction
\beq
\lambda_{ABCD} \leq \lambda_{\rm diag} \leq  \frac{2\pi}{\beta},
\eeq
where $\lambda_{\rm diag} \equiv {\rm min} (\lambda_{ABAB},\lambda_{ADAD},\lambda_{CBCB}, \lambda_{CDCD})$. Namely, if the rate of growth are different for each term, we can say that $\lambda_{ABCD}$ is smaller than the minimum of $\lambda_{ABAB}$, $\lambda_{CBCB}$, etc, which are all smaller than $2\pi/\beta$ by the chaos bound. 

Similarly to the previous case, we can assume all OTOC have the same rate of growth, and then we can also bound the amplitude $\varepsilon$ of the growing piece. The bound we obtain from the previous analysis, equation \eqref{eq:boundabac}, is 
\bea\label{gencond}
&&\hspace{-0.5cm}(\alpha_1^2 \varepsilon_{ABAB} +\alpha_2^2 \varepsilon_{ADAD} + 2 \alpha_1 \alpha_2 \varepsilon_{ABAD} )(\alpha_1^2 \varepsilon_{CBCB} +\alpha_2^2 \varepsilon_{CDCD} + 2 \alpha_1 \alpha_2 \varepsilon_{CBCD} )\nn
&&~~~-(\alpha_1^2 \varepsilon_{ABCB} +\alpha_2^2 \varepsilon_{ADCD} + 2 \alpha_1 \alpha_2 \varepsilon_{ABCD} )^2\geq 0,
\ea 
which should be satisfied for any choice of $\alpha_1$, $\alpha_2$. Since this condition is invariant under a rescaling of $\alpha_i \to \lambda \alpha_i$, we can fix $\alpha_1=1$. Then this condition \eqref{gencond} is equivalent to the positivity of a quartic polynomial on the variable $\alpha_2$ with coefficients depending on the $\varepsilon$'s. \footnote{For a general quartic polynomial $P(x)=a x^4 + b x^3 + c x^2 + d x + e $ it should have $a,e>0$ and the condition having four complex roots is to have a positive discriminant $\Delta(P)\geq0$, and a positive $8 ac-3b^2\geq0$.} When these conditions are written in terms of the amplitudes $\varepsilon$'s they look algebraically complicated and not very enlightening. 

To simplify the discussion we can use the previous bound $\varepsilon_{ABAD}^2 \leq \varepsilon_{ABAB} \varepsilon_{ADAD}$ and $ \varepsilon_{CBCD}^2 \leq \varepsilon_{CBCB} \varepsilon_{CDCD}$ to complete the square in the first line of equation \eqref{gencond}. Then we can derive a non-optimal bound on the most generic $\varepsilon_{ABCD}$ as 
\beq
\varepsilon_{ABCD} ^2 \leq  4( \sqrt{ \varepsilon_{ADAD}\varepsilon_{CBCB}} +\sqrt{ \varepsilon_{ABAB}\varepsilon_{CDCD}})^2.
\eeq
Even though this bound is not optimal it shows in a more transparent way how the prefactor of the off-diagonal OTOC is bounded by the diagonal ones. This is the main conceptual point, in a holographic setting this shows hows a generic interaction is bounded by the gravitational interactions.

As a final comment,we can also consider a correlator of the type  
\beq
F(t)= {\rm Tr} [ y V y W(t) y V y W(t) ]  ,~~~ V=\alpha_1 A + \alpha_2 C,~~{\rm and}~~W=\alpha_3 B + \alpha_4 D
\eeq 
One might wonder whether the $\varepsilon>0$ constraint for $F(t)$ by varying all $\alpha$'s independently we can derive a bound on $\varepsilon_{ABCD}$ stronger than the one above which was derived by steps. It is easy to see that this is not the case, and considering the most general linear combination does not generate new constraints compared to equation \eqref{gencond}.

\subsection{Non-Hermitian Operators}\label{sec:nonh}
So far we have discussed OTOC between arbitrary hermitian operators. Some of the steps for the bound on chaos argument from \cite{MSS} do not directly work for non-hermitian operators. We will show here that the bound on hermitian operators is enough to prove this generalization.   

Consider the following OTOC between general non-hermitian operators
\beq\label{eq:vdwdvw}
F(t)= {\rm Tr} [ y V^\dagger y W^\dagger(t) y V y W(t) ]  
\eeq
We will show how the bound on ${\rm Re}~F(t)$ derives from the bound on hermitian operators. This quantity is related to the double commutator between non-hermitian $V$ and $W$ that appear in the definition of chaos.

We will expand a general non-hermitian operator $\mathcal{O}$ in two hermitian components $\mathcal{O}_R = (\mathcal{O}+\mathcal{O}^\dagger)/2$ and $\mathcal{O}_I = (\mathcal{O}-\mathcal{O}^\dagger)/(2i)$, for $\mathcal{O}=V,W$. To simplify the expressions we write below, we will use a shorthand for the OTOC defining $\lb ABCD\rb \equiv {\rm Tr} [ y A(0) y B(t) y C(0) y D(t)]$. Starting from the right hand side of \eqref{eq:vdwdvw}, expanding and using the cyclic property of the trace it is easy to show that  
\bea\label{eq:nonhermtotl}
{\rm Re} ~F(t)&=&{\rm Re}~{\rm Tr} [ y (V_R-i V_I) y W^\dagger(t) y (V_R+i V_I) y W(t) ]  \nn
&=&{\rm Re}~[ \lb V_R W^\dagger V_R W\rb + \lb V_I W^\dagger V_I W\rb- i (\lb V_I W^\dagger V_R W\rb-\lb V_R W^\dagger V_I W\rb)]\nn
&=&\lb V_R W^\dagger V_R W\rb + \lb V_I W^\dagger V_I W\rb 
\ea
where we used that $ \lb V_I W^\dagger V_R W\rb^* = \lb W^\dagger V_R W V_I \rb = \lb V_I W^\dagger V_R W\rb$ and similarly for $\lb V_R W^\dagger V_I W\rb$, implying they are both real, and therefore the last term in the right hand side of the equation above is purely imaginary. 

Now we can expand $W$ and use that $\lb A B C B\rb$ is real for hermitian operators, to show
\beq\label{eqnonh}
{\rm Re} ~F(t) = \lb V_R W_R V_R W_R \rb + \lb V_R W_I V_R W_I\rb +\lb V_I W_R V_I W_R \rb + \lb V_I W_I V_I W_I\rb,
\eeq
Then if we write ${\rm Re} ~F(t) = F_d - \varepsilon ~e^{\lambda t}$, the chaos bound on the growth rate automatically applies to each term individually on the right hand side implying $\lambda \leq 2\pi/\beta$. Moreover since all terms appear with a plus sign, the bound on the sign of the prefactor still applies, implying $\varepsilon\geq0$.

Taking the usual chaos bound for non-hermitian operators as a starting point we can derive analogous results as in the previous section for general non-hermitian operators.

\section{An Example: 2d CFTs}\label{sec:2dCFT}

In the context of 2d CFTs one can show that at large $c$, a large gap in the twist is enough to obtain maximal chaos by using results from semiclassical limits of Virasoro conformal blocks \cite{Roberts:2014ifa} \footnote{Other studies of chaos in 2d CFT from different perspectives can be found in \cite{Jackson:2014nla, Turiaci:2016cvo} (see also \cite{Perlmutter:2016pkf}). }. This is given purely by a product of stress tensors acting on the identity and therefore can be interpreted as coming in the bulk from a purely gravitational interaction.

In this section we want to study in a simple setup which role the inelastic version of the chaos bound plays for large $c$ 2d CFTs with a sparse spectrum. Under some assumptions, we will study the general behavior of off-diagonal OTOC. We propose that a resummation of intermediate channels can be written as a single non-vacuum block corresponding to an operator with an effective dimension and an effective spin. 

From a bulk perspective this constrains matter interaction (OPE coefficients of arbitrary operators) using the chaos bound.  

\subsection{Kinematics} 
In any 2d CFT an arbitrary four point function can be written as  
\beq\label{eq:4pftnction}
\lb W_1(z_1,\bar{z}_1)W_2 (z_2,\bar{z}_2) V_3 (z_3, \bar{z}_3) V_4 (z_4,\bar{z}_4) \rb = \frac{1}{z_{12}^{h_1+h_2} z_{34}^{h_3+h_4} }\frac{1}{\bar{z}_{12}^{h_1+h_2} \bar{z}_{34}^{h_3+h_4} } G(z,\bar{z})
\eeq
where $G(z,\bar{z})$ can be expanded in Virasoro conformal blocks and the cross-ratio is defined as $z=\frac{z_{12}z_{34}}{z_{13}z_{24}}$ and a similar anti-holomorphic version. The operators are arbitrary but we use the letters $V$ and $W$ to indicate which ones will be at time $0$ ($V$'s) and which at time $t$ ($W$'s). Schematically we expand the four-point function as
\beq\label{eq:OPEdef}
G(z,\bar{z}) = \sum_{p} C_{12p}C_{34p}~ \mathcal{F} \big[{}^{h_1}_{h_2}{}^{h_3}_{h_4}\big](h_p, z)~ \mathcal{F}\big[{}^{\bar{h}_1}_{\bar{h}_2}{}^{\bar{h}_3}_{\bar{h}_4}\big](\bar{h}_p,\bar{z})
\eeq 
where $\mathcal{F}_{h_p} \big[{}^{h_1}_{h_2}{}^{h_3}_{h_4}\big](z)$ are the Virasoro blocks corresponding to a Virasoro primary operator $\mathcal{O}_p$ with (anti)holomorphic weights $h_p$($\bar{h}_p$), dimension $\Delta = h_p + \bar{h}_p$ and spin $s=|h_p-\bar{h}_p|$. The blocks are normalized such that $\mathcal{F}_h(z) = z^h(1+\ldots)$ for a small $z$ expansion.

In going from Euclidean to Lorenzian signature, different orders of operators are encoded in the monodromy of the blocks \cite{Rehren:1987ur}. Following \cite{Roberts:2014ifa} we choose the kinematics of the correlator to be 
\bea
&&z_1 = e^{\frac{2\pi}{\beta} ( t+i\epsilon_1)},~~~~\hspace{0.08cm}z_2 = e^{\frac{2\pi}{\beta} ( t+i\epsilon_2)},~~~~\hspace{0.08cm}z_3 = e^{\frac{2\pi}{\beta} ( x+i\epsilon_3)},~~~z_4 = e^{\frac{2\pi}{\beta} ( x+i\epsilon_4)},\nn
&&\bar{z}_1 = e^{-\frac{2\pi}{\beta} ( t+i\epsilon_1)},~~~\bar{z}_2 = e^{-\frac{2\pi}{\beta} ( t+i\epsilon_2)},~~~\bar{z}_3 = e^{\frac{2\pi}{\beta} ( x-i\epsilon_3)},~~~\bar{z}_4 = e^{\frac{2\pi}{\beta} ( x-i\epsilon_4)}
\ea
where $\epsilon_1<\epsilon_3<\epsilon_2<\epsilon_4$ and as we raise the time from $0$ to $t$ the cross-ratio $z$ goes once around $z=1$ while $\bar{z}$ remains in the first sheet. For times larger than the dissipation time, which is of order $\beta$, we can evaluate the blocks on the second sheet with cross-ratios $z \approx - \epsilon_{12}^\star \epsilon_{34} e^{\frac{2\pi}{\beta}(x-t)}$ and $\bar{z} \approx - \epsilon_{12}^\star \epsilon_{34} e^{-\frac{2\pi}{\beta}(x+t)}$, where $\epsilon_{ij} = i (e^{\frac{2\pi}{\beta} i \epsilon_i} - e^{\frac{2\pi}{\beta} i \epsilon_j})$. For a configuration with operators equally spaced on the thermal circle, $\epsilon_{12}^\star \epsilon_{34}=4i$ and $z = -4 i  e^{\frac{2\pi}{\beta}(x-t)}$ and $\bar{z} = - 4i e^{-\frac{2\pi}{\beta}(x+t)}$.

If we normalize operators by their 2pt function on the plane $\lb O (x) O(0) \rb =(x)^{-2\Delta_O}$, then each term in the factorized answer for the four-point function gives a factor of ${\rm Tr} [ y^2 O y^2 O] = (\pi/\beta )^{2\Delta_O}$. This also appears from the position dependent prefactor in the right hand side of equation \eqref{eq:4pftnction} when the operators are all different.

\subsection{Elastic case} 

In this case we take the two operators at $t=0$ and $t$ to be the same $V_3 = V_4 = V$ and $W_1 = W_2 = W$. Then the identity appears on the intermediate channel in the OPE written above in equation \eqref{eq:OPEdef}. Assuming a large twist gap we can approximate the full correlator by the vacuum block. Following \cite{Roberts:2014ifa} we will take a large $c$ limit with $h_1/c\ll1$ fixed but small and $h_3\gg 1$. Then we can make use of heavy-light semiclassical blocks found in \cite{Fitzpatrick:2014vua}. The final answer is given by
\bea\label{eq:vacuumblockapprox}
\frac{ {\rm Tr} [ y V y W(t) y V y W(t)]}{ {\rm Tr}[y^2 V y^2 V]{\rm Tr}[y^2 W y^2 W]} &\approx& \big( 1+\frac{6 \pi  h_1}{c} e^{\frac{2 \pi}{\beta}(t-x)}  \big)^{-2h_3}\nn
&\approx&1 - \frac{12 \pi  h_1  h_3}{c}  e^{\frac{2 \pi}{\beta}(t-x)} + \ldots
\ea
which saturates the chaos bound. In the first line we wrote the chaos limit of the identity Virasoro block, while in the second line we focus on times $\beta^{-1} \ll t \ll t_{\rm sc}=\frac{\beta}{2\pi} \log c$. This was formally done at infinite gap. Corrections from finite gap and how the correlator Reggeize was recently studied in reference  \cite{Chang:2018nzm}. Within this approximation, for times larger than the scrambling time $t\gg t_{\rm sc} $ this OTOC goes to zero exponentially at a rate related by the quasinormal modes.

\subsection{Inelastic case}
In the inelastic case we can take four arbitrary operators $W_1$, $W_2$, $V_3$ and $V_4$. In order to have analytic control over the Virasoro conformal blocks we will take large $c$ with $1 \ll h_3,h_4$ and fixed $h_1/c$, $h_2/c$ but small. Moreover we also take $h_{12}=(h_1-h_2)/2$ and $h_{34}=(h_3 -h_4)/2$ to be of order one such that the results of  \cite{Fitzpatrick:2014vua} apply. 

We can take a basis of operators such that the identity block does not appear on the intermediate channel (this is automatic if the dimensions are different). Instead within a similar approximation as in the elastic case, we need to consider light intermediate operators of low twist. The semiclassical Virasoro block was also computed in this case, when the channel dimension $h_p$ is of order one \cite{Fitzpatrick:2014vua}. After going to the second sheet and using the chaos kinematics we get 
\beq\label{eq:nonvacblock}
\mathcal{F} \big[{}^{h_1}_{h_2}{}^{h_3}_{h_4}\big](h, z) = \left(\frac{1}{ 1-\frac{12 \pi i  (h_1+h_2)}{cz}  }\right)^{h_3+h_4-h} z^h {}_2 F_1 (h - h_{12}, h+ h_{34}, 2h, z)|_{{\rm 2nd~sheet}} ,
\eeq
where the hypergeometric function comes from $SL(2)$ descendants and it is evaluated on the second sheet. 

For times between dissipation and scrambling, the first term in the right hand side of \eqref{eq:nonvacblock} is constant and the exponential growth comes from the hypergeometric function 
\beq
{\rm Tr} [ y V_3 y W_1(t) y V_4 y W_2(t)] = N_\beta \sum_{p} C_{12p}C_{34p}~d_p~e^{ \frac{2\pi}{\beta}(s_p-1)t}e^{- \frac{2\pi}{\beta}(\Delta_p-1) x},
\eeq
where $\Delta_p$, $s_p$ is the dimension and spin of the intermediate channel operator $O_p$. The normalization coming from the prefactor of \eqref{eq:4pftnction} is $N_\beta = (\pi/\beta)^{\Delta_1 +\Delta_2 + \Delta_3 + \Delta_4}$. $d_p$ is a coefficient 
%\beq
%d_p = \frac{8\pi ~e^{i \pi (h_{34}-h_{12})}}{(i4)^{h_p-\bar{h}_p }(2h-1)}\frac{\Gamma(2h)}{\Gamma(h\pm h_{12})}\frac{\Gamma(2h)}{\Gamma(h\pm h_{34})},
%\eeq
\beq
d_p = \frac{8\pi }{(i4)^{h_p-\bar{h}_p }(2h-1)}\frac{\Gamma(2h_p)}{\Gamma(h_p\pm h_{12})}\frac{\Gamma(2h_p)}{\Gamma(h_p\pm h_{34})},
\eeq
where $\Gamma(a\pm b) =\Gamma(a+b)\Gamma(a-b)$. This factor depends on the dimension of the intermediate channel and comes from the evaluation of the hypergeometric function in \eqref{eq:nonvacblock} on the second sheet.

The growing part of elastic OTOC for holographic CFTs are dominated by the vacuum block, dual to gravitational interactions in the bulk. The growing part of inelastic OTOCs is not related to gravitational interactions. The fact that Virasoro descendants are irrelevant for the calculation of inelastic OTOCs is a manifestation of this fact. This is not completely obvious and we give the details in the appendix. The amplitude of the growing piece of the inelastic OTOC is small due to the fact that the OPE coefficients are subleading in $N$ and in the gap. 

Lets first assume that the spins are bounded. Then by looking at this expression assuming a large twist gap we can conclude that in the low-lying part of the spectrum all particles must have spin $s<2$. If a primary happens to have $s=2$ then its interactions with other particles cannot be stronger than gravity (bounds OPE coefficient). 

If there happens to be a light particle with $s>2$, then its contribution should be Reggeize among the low-lying spectrum to give an effective spin $s_{\rm eff} <2$. Then we can see the statement in the previous paragraph as a statement about inelastic Pomerons in the theory.  
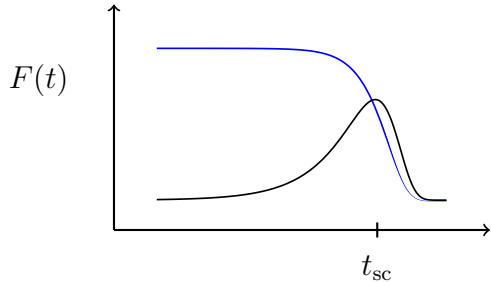
\begin{figure}[t!]
\begin{center}
\begin{tikzpicture}[scale=1]
\node[inner sep=0pt] (russell) at (2.5,1.5)
    {\includegraphics[width=.25\textwidth]{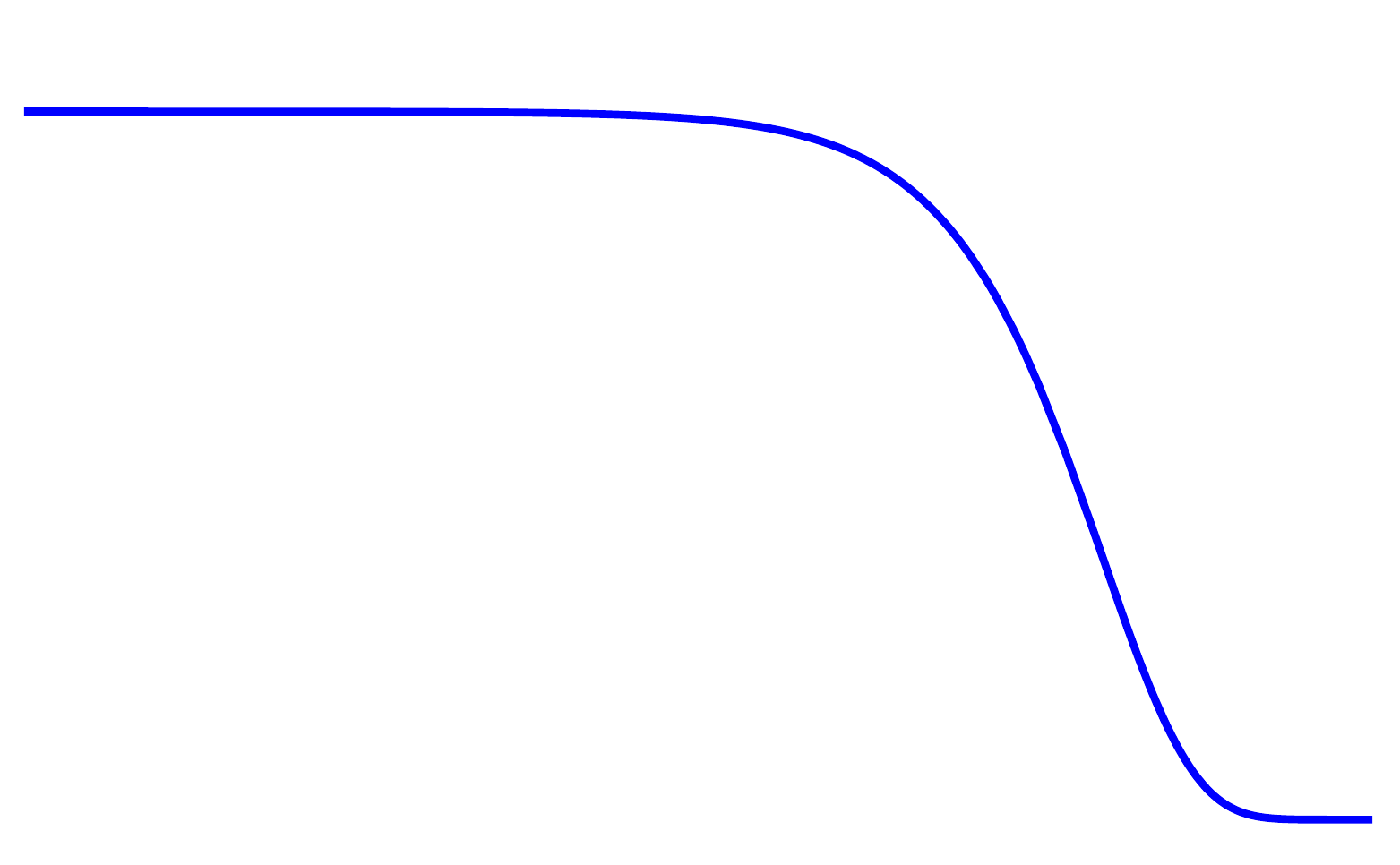}};
    \node[inner sep=0pt] (russell) at (2.5,1.5)
    {\includegraphics[width=.25\textwidth]{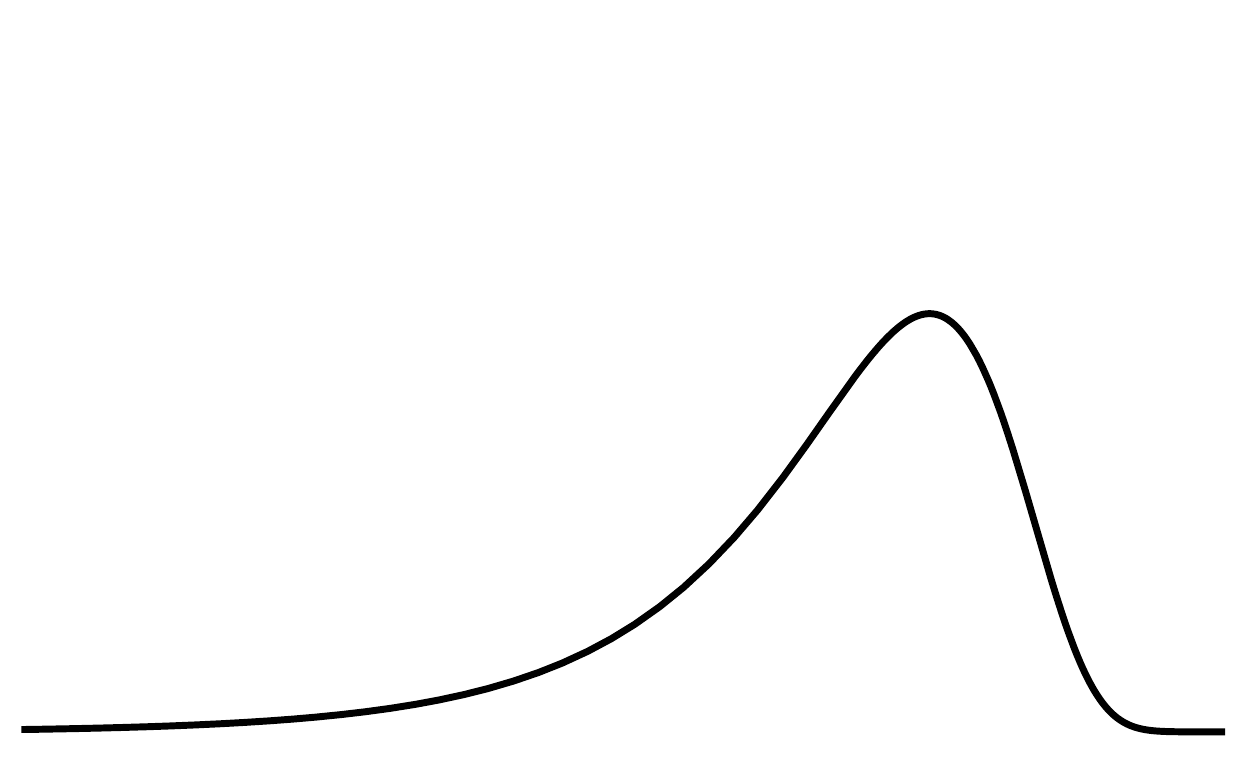}};
\draw[thick, ->] (0,0) -- (5,0);
\draw[thick, ->] (0,0) -- (0,3);
\draw[thick] (3.5,0-0.1)--(3.5,0+0.1);
\draw (3.5,-0.5) node {$t_{\rm sc}$};
\draw (-1,2) node {$F(t)$};
    \end{tikzpicture}
    \end{center}
    \vspace{-0.85cm}    
    \caption{\small  Sketch of a typical behavior of inelastic OTOC $F(t)$ in 2d CFT (black curve) as a function of time, assuming its approximated by a non-vacuum block with effective dimension $\Delta_{\rm eff}$ and spin $s_{\rm eff}$. Initially the OTOC grows exponentially with rate $\lambda = \frac{2\pi}{\beta} (s_{\rm eff}-1)$. For late times the fast decay is controlled by the quasi-normal modes. In blue we show a typical elastic OTOC. }
    \label{fig:temp-check}
    \end{figure}

The calculation required in the previous paragraph to evaluate the inelastic OTOC is complicated, even in the case of elastic OTOC \cite{Chang:2018nzm}. We can conjecture that the result of summing over infinite spins is equivalent to a single non-vacuum block with effective $h_{\rm eff}$, $\bar{h}_{\rm eff}$ and effective dimension $\Delta_{\rm eff}=h_{\rm eff} + \bar{h}_{\rm eff}$ and spin $s_{\rm eff}=|h_{\rm eff} - \bar{h}_{\rm eff}|$. With this assumption the inelastic OTOC is given by 
\beq
N^{-1}_\beta~{\rm Tr} [ y V_3 y W_1(t) y V_4 y W_2(t)] = ~C_{12}C_{34}~\mathcal{F} \big[{}^{h_1}_{h_2}{}^{h_3}_{h_4}\big](h_{\rm eff}, z) ~\bar{z}^{\bar{h}_{\rm eff}},
\eeq
where the holomorphic block is given by equation \eqref{eq:nonvacblock}, and $C_{12}$ ($C_{34}$) are effective couplings between operators $W_1$, $W_2$ ($V_3$, $V_4$) and the effective Pomeron state $h_{\rm eff}, \bar{h}_{\rm eff}$. Depending on the effective spin, $C_{12}C_{34}$ might be bounded by the dimensions of the external operators, following \eqref{ec:introbound}.

With the proposal of the previous paragraph, we can analyze times longer than scrambling $t_{\rm sc} \lesssim t$. In this case the situation changes and the part of the block coming from the Virasoro descendants dominate. Namely, the first factor in the right hand side of \eqref{eq:nonvacblock} decays exponentially. Assuming the behavior of the correlator is equivalent to a single block of dimension $\Delta_{\rm eff}$ of order one and spin $s_{\rm eff}<2$, the OTOC decays exponentially as $\sim e^{- \frac{2\pi}{\beta} (h_3+h_4)t} $. Under the assumption of the previous paragraph, we show in figure \ref{fig:temp-check} the behavior of a typical OTOC between different operators. 
%\beq
%{\rm Tr} [ y V_3 y W_1(t) y V_4 y W_2(t)] \sim e^{- \frac{2\pi}{\beta} (h_3+h_4)t} 
%\eeq

To summarize, inelastic OTOCs grow exponentially in a way controlled by a Pomeron exchange (unrelated to gravity), until the scrambling time $t_{\rm sc}$ at which the correlator begins to decay according to the quasinormal modes frequencies. This is expected since this decay is due to the bulk-boundary propagators appearing in the bulk calculation of the OTOC. The picture and the proposal that emerges from this analysis should be worked out in more detail following \cite{Chang:2018nzm} and \cite{Collier:2018exn} (see also \cite{Liu:2018iki} and \cite{Hampapura:2018otw}), but we leave it for future work.

\section{Open Questions}\label{Sec:conc}

To conclude we would like to state some open questions. 

It would be nice to compute the off-diagonals OTOC introduced here for SYK models \footnote{In particular, the approach of \cite{Berkooz:2018jqr} might be very useful for this.}. Following the notation of \cite{Kitaev:2017awl} and \cite{Gu:2018jsv}, we can write a generic OTOC as a convolution between form factors describing the coupling of external operators to an effective `scramblon' mode, and the scramblon propagator which grows exponentially with time. In this perspective the bound stated here constraints the behavior of the general form factors appearing in these models. Their rate of growth in time is bounded by the growth of the scramblon mode through the elastic OTOC. 

Moreover, the mode responsible to the Lyapunov behavior of inelastic OTOC might not be the same as the one for elastic case (for example, it might not have the quantum numbers of the vacuum). Therefore there must be a mode which similarly to the Schwarzian mode generates exponential growth\footnote{From the perspective of \cite{MTV} the problem is analogous to finding a generalization of Liouville theory that allows primary operators with spin.}. A simple proposal would be a similar mode living on ${\rm Diff}(S^1)/U(1)$ instead of ${\rm Diff}(S^1)/SL(2)$, but this requires further study.   

This question can be extended to higher dimensions. In 2d CFTs the maximal chaos behavior coming from the identity block can be understood as coming from a Goldstone mode of broken reparametrization invariance \cite{Turiaci:2016cvo}. It would be nice to find a description of a similar soft mode producing exponential growth of off-diagonal OTOC, related to non-vacuum representations. To analyze this problem 2d versions of SYK might be useful as explicit examples \cite{Gu:2016oyy, Turiaci:2017zwd, Murugan:2017eto, Berkooz:2017efq}. In the particular case of 2d CFTs it would be interesting to repeat the analysis of \cite{Chang:2018nzm} for a general OTOC.

The analysis of this paper can be extended to higher order OTOC with more than four operators. The diagonal version of these correlators was studied in \cite{Haehl:2017pak} (see also \cite{Haehl:2018izb}). Results in this direction were derived in \cite{Basu:2018akv}.

Finally, after understanding how correlators Reggeize in the chaos limit, it would be nice to recast the bound derived in this paper as a bound on OPE data. This might also help sharpen the statement about gravity being the highest spin, strongest, interaction. 

\bigskip

\begin{center}
{\bf Acknowledgements}
\end{center}

\vspace{-2mm}

We want to thank G. Horowitz, J. Maldacena, M. Rangamani and D. Stanford for discussions and comments on the draft. This work was supported by a Fundamental Physics Fellowship. We also benefited from the workshop ``Chaos
and Order: From strongly correlated systems to black holes" at KITP, supported in part by the National Science Foundation under Grant No. NSF PHY-1748958 to the KITP.

\begin{appendix}
\section{Semiclassical Virasoro Blocks}\label{app:blocks}
\end{appendix}
In this paper we used a Virasoro conformal block with two external light operators of weights $h_L\pm \delta_L$ and two external heavy operators of weights $h_H \pm \delta_H$. In the large $c$ limit with $h_H/c$ fixed these were obtained in \cite{Fitzpatrick:2014vua} for a light intermediate channel $h$. Within this approximation it is given by 
\beq\label{app:fitzkapblock}
\mathcal{F} \big[{}^{H}_{H}{}^{L}_{L}\big](h, z)= (1-w)^{(h_L+\delta_L)(1-\alpha^{-1})}\left( \frac{w}{\alpha ~z} \right)^{h-2h_L} z^h {}_2 F_1 \left( h - \frac{\delta_H}{\alpha} , h+\delta_L , 2 h , w \right)
\eeq
where $\alpha=\sqrt{1-24 h_H/c}$ and $w(z)=1-(1-z)^\alpha$. In the case of pairwise identical operators, $\delta_H=\delta_L=0$, and for the vacuum channel, this expression gives 
\beq
\mathcal{F} \big[{}^{H}_{H}{}^{L}_{L}\big](0, z) =  (1-w)^{(h_L+\delta_L)(1-\alpha^{-1})}\left( \frac{w}{\alpha~z} \right)^{-2h_L}.
\eeq
In the limit for which $h_H/c \ll 1$ ($\alpha\approx 1-12h_H/c$), for small $z$ in the second sheet the block is approximately $ \mathcal{F} \big[{}^{H}_{H}{}^{L}_{L}\big](0, z) \approx \left(\frac{z}{1-(1-z)^\alpha} \right)^{2h_L}$. This expression reproduces equation \eqref{eq:vacuumblockapprox}, which is the main result in \cite{Roberts:2014ifa}. 

For the case of generic intermediate channel we can use the general formula, go to the second sheet and evaluate in the chaos limit. This gives the same answer as doing the analytic continuation of the expression 
\beq\label{app:eq3}
\mathcal{F} \big[{}^{H}_{H}{}^{L}_{L}\big](h, z)= \left( \frac{z}{1-(1-z)^\alpha} \right)^{2h_L-h} z^h {}_2 F_1 \left( h - \frac{\delta_H}{\alpha} , h+\delta_L , 2 h , w \right)
\eeq
The first term is equivalent to the vacuum block with shifted dimensions. This term is entirely due to Virasoro descendants. The evaluation of this term in the second sheet is therefore 
\beq\label{eq:prefactorapp}
\left( \frac{z}{1-(1-z)^\alpha} \right)^{2h_L-h} \Bigg|_{2{\rm nd~sheet}} \approx \left( \frac{1}{1-\frac{24 \pi i h_H}{cz} } \right)^{2h_L -h}
\eeq
In the chaos limit, and for times smaller than the scrambling time $c z \gg 1$ ($t\ll \frac{\beta}{2\pi} \log c$), this gives approximately a constant $1$. After the scrambling time $c z \ll 1$, this term decays and controls the decay of correlators. The second term of \eqref{app:eq3} evaluated on the second sheet and for small $z$ gives 
\beq
z^h{}_2 F_1 (h - \frac{h_{H}}{\alpha}, h+ h_{L}, 2h, w)|_{{\rm 2nd~sheet}}=  \frac{2\pi i ~e^{i \pi (h_{L}-h_{H})}}{2h-1}\frac{\Gamma(2h)}{\Gamma(h\pm \frac{h_{H}}{\alpha})}\frac{\Gamma(2h)}{\Gamma(h\pm h_{L})} \frac{z}{(\alpha z)^h}
\eeq
For $h_H/c$ with $\alpha \sim 1$ this gives the same as the evaluation of the global block $z^h{}_2 F_1 (h - h_{H}, h+ h_{L}, 2h, z)$. Therefore we can approximate the Virasoro block by 
\beq
\mathcal{F} \big[{}^{H}_{H}{}^{L}_{L}\big](h, z)\approx \left( \frac{z}{1-(1-z)^{1-12 h_H/c}} \right)^{2h_L-h} z^h {}_2 F_1 \left( h - \delta_H , h+\delta_L , 2 h , z \right).
\eeq
The evaluation of this expression above in the second sheet for small $z$ gives the same answer as applied on the original \eqref{app:fitzkapblock}. Therefore all the effects of Virasoro descendants in the chaos limit come from the prefactor in the equation above.

\begingroup\raggedright\endgroup

%\bibliography{trapbib}
\end{document}